\documentclass{aa}
\usepackage{graphicx}
\usepackage{txfonts}
\begin{document}

\title{Antenna Instrumental Polarization 
and its Effects on $E$- and $B$-Modes for CMBP
Observations
}
\author{Ettore~Carretti\inst{1} \and
        Stefano~Cortiglioni\inst{1} \and
        Carla~Sbarra\inst{1} \and
        Riccardo~Tascone\inst{2}}

\institute{I.A.S.F/C.N.R. Bologna, Via Gobetti 101, I-40129 Bologna, Italy.
           \and
           I.E.I.I.T./C.N.R. Torino, C.so Duca degli Abruzzi 24,
           I-10129 Torino, Italy.
          }

\offprints{Ettore Carretti, \email{carretti@bo.iasf.cnr.it}}
%------------------------------------------------------------------------------

\abstract{We analyze the instrumental polarization
generated by the antenna system (optics and feed horn)
due to the unpolarized sky emission.
Our equations
show that it is given by the convolution of the unpolarized emission map
$T_b(\theta, \phi)$ with a sort of {\it instrumental polarization beam}
$\Pi$ defined by the co- and cross-polar patterns of the antenna.
This result is general, it
can be applied to all antenna systems
and is valid
for all schemes to detect polarization, like correlation and differential
polarimeters. The axisymmetric case is attractive: it generates an
$E$-mode--like
$\Pi$ pattern, the contamination does not depend on the scanning strategy
and the instrumental polarization map does not
have $B$-mode contamination, making axisymmetric systems suitable
to detect the faint $B$-mode signal of the
Cosmic Microwave Background Polarization.
The $E$-mode of the contamination only affects the
FWHM scales leaving the larger ones significantly cleaner.
Our analysis is also applied to the SPOrt experiment where we find that
the contamination of the $E$-mode is negligible in the $\ell$-range of
interest for CMBP large angular scale
investigations (multipole $\ell < 10$).
}

\date{Received / Accepted}

\titlerunning{Antenna Instrumental Polarization}
\authorrunning{E. Carretti et al.}
\maketitle

\section{Introduction}

The Polarization of the Cosmic Microwave Background (CMBP) represents
a powerful tool to determine cosmological parameters, investigate the
epoch of the formation of the first galaxies and have
an insight into the inflationary era (e.g. see
Zaldarriaga, Spergel \& Seljak 1997,
Kamionkowski \& Kosowsky 1998).

In spite of its importance the CMBP signal
is weak: The $E$-mode component is about 1-10\% of the already faint
CMB anisotropy,
depending on the angular scale; The $B$-mode is even
weaker, with an emission level that can be
more than 3 orders of magnitude
below the anisotropy,
depending on the tensor-to-scalar perturbation ratio $T/S$
(e.g. see Kamionkowski \& Kosowsky 1998).

The investigation of CMBP
is at the beginning with a first detection of the $E$-mode
claimed by the DASI team (Kovac et al. 2002)
and a first measurement of the TE cross-spectrum provided
by the WMAP team (Kogut et al. 2003). Anyway, we are far
from a full characterization of the $E$-mode
and the $B$-mode is still elusive.

In this frame it is important to deal with instruments
with low contamination, expecially in the $B$ component,
for which even a 0.1\% leakage from the Temperature
anisotropy might wash out any detection.
One should also note that the contamination from the anisotropy term
is not easily removed by destriping techniques, its level
being not constant over the scans
performed to observe the sky (e.g. see
Revenu et al. 2000, Sbarra et al. 2003 and references
therein for descriptions of destriping techniques), so that
a clean instrument is necessary to avoid
complex data reductions.

A clear understanding of the contamination
from the unpolarized background introduced by
the instrument itself is thus mandatory to properly design instruments,
as confirmed by the presence of
several works on this subject (e.g.
Carretti et al. 2001, Leahy et al. 2002, Kaplan \& Delabrouille 2002,
Franco et al. 2003).

In this paper we analyze the instrumental polarization generated
in the antenna system (optics and feed horn)
by the unpolarized emission.
We provide the equations to compute the contamination
in both the $Q$ and $U$ outputs as a  function
of the antenna properties. In particular, we find
that a sort of {\it instrumental polarization beam} $\Pi$
is acting and that the contamination is the result of the convolution
between this beam and the unpolarized emission field.

Our results are general
and can be applied independently of the
scheme implemented to measure the linear Stokes parameters $Q$ and $U$.
For instance, they can be applied to both correlation and differential
polarimeters.

Finally, we analyze the contamination in the $E$ and $B$-mode power spectra.
We find that axisymmetric systems do not contaminate
the $B$-mode making such optics a suitable solution
for the detection of this faint signal.
In particular, we analyze the
SPOrt\footnote{http:$\backslash$$\backslash$sport.bo.iasf.cnr.it}
experiment, for which we study
the instrumental polarization beam and compare its contamination
in the $E$ and $B$ mode to the CMBP signal.

The paper layout is as follows: In Section~\ref{instrPolSec}
we present the equations to compute the instrumental polarization due to the
antenna
in the general case; In Section~\ref{axSymmSec} we analyze the special
case of axisymmetric systems; In Section~\ref{EBSec} we study the
effects on the $E$ and $B$ mode power spectra for axisymmetric systems; 
Finally, Section~\ref{conclusions} gives the conclusions.

\section{Instrumental Polarization of the Antenna System}\label{instrPolSec}

The instrumental polarization generated by the antenna system can
be estimated by its contamination into the two linear Stokes
parameters $Q$ and $U$. These quantities can be written as the
correlation-product between the Left and the Right-handed circular
polarizations gathered by the antenna. In terms of spectral distribution
the Stokes parameters write 
\begin{equation}
    Q+jU =  b_R \, b_L^*, \label{queq}
\end{equation}
where $b_{L}$ and $b_R$ are the waveguide power waves received
from the antenna system and corresponding to the left and the
right polarizations, respectively,
the  symbol $^*$ denotes the complex conjugate and $j$ the
complex unit. With reference to the radiation characteristics  of
the antenna, the two power waves can be expressed as the
integrals onto the 2D--sphere
\begin{eqnarray}
  b_L &=& \frac{\lambda}{\sqrt{4 \pi Z_0}}\int_\Omega  {\bf E} \cdot {\bf h}_L \,d \Omega, \nonumber\\
  b_R &=& \frac{\lambda}{\sqrt{4 \pi Z_0}} \int_\Omega  {\bf E} \cdot {\bf h}_R \,d \Omega, \label{blbreq}
\end{eqnarray}
where ${\bf E}$ is the electric field spectral distribution of the
incoming radiation, $\lambda$ is the free-space
wavelength, $Z_0=\sqrt{\mu_0/\epsilon_0}$  is the free-space
impedance and ${\bf h}_L$ and ${\bf h}_R$  are the complex
vector radiation patterns corresponding to the left and the right
circular polarization channels, respectively, and whose square
magnitudes give the antenna gains.

The instrumental polarization is defined as the spurious outputs
$Q_{\rm sp}$ and $U_{\rm sp}$ detected  by the instrument in the
case of unpolarized radiation. Under this condition, the
correlation product between orthogonal components of the incoming
radiation is zero, hence, from equations~(\ref{queq})
and~(\ref{blbreq}), we have
\begin{equation}
    Q_{\rm sp} + j\,U_{\rm sp} =
       {\lambda^2\over 4 \pi Z_0 } \int_\Omega
       {|{\bf E}|^2 \over 2}\; {\bf h}_R \cdot {\bf h}_L^*
       \;d\Omega.
       \label{SpCirc}
\end{equation}

The $Q$ and $U$ Stokes parameters can be also evaluated by means
of  the difference and the correlation of the two linear
polarizations  gathered by the antenna.  
Let us consider the two power
waves received by the antenna system related to the two linear
polarizations
\begin{eqnarray}
  b_X &=& \frac{\lambda}{\sqrt{4 \pi Z_0}}\int_\Omega  {\bf E} \cdot {\bf h}_X \,d \Omega, \nonumber\\
  b_Y &=& \frac{\lambda}{\sqrt{4 \pi Z_0}} \int_\Omega  {\bf E} \cdot {\bf h}_Y \,d \Omega, \label{bLin}
\end{eqnarray}
where ${\bf h}_X$ and ${\bf h}_Y$  are the complex vector
radiation patterns corresponding to the $X$ and the $Y$ linear
polarization channels, respectively.
Reminding (e.g. see Kraus 1986)
\begin{eqnarray}
    Q &=& \frac{1}{2}\left[|b_X|^2 - |b_Y|^2\right], \nonumber\\
    U &=& \Re (b_X\; b_Y^*),
    \label{quLin}
\end{eqnarray}
the detected spurious outputs $Q_{\rm sp}$ and $U_{\rm sp}$ are:
\begin{eqnarray}
    Q_{\rm sp} &=&
       {\lambda^2\over 4 \pi Z_0 } \int_\Omega
       \frac{|{\bf E}|^2}{2} \; \left[\frac{1}{2}(|{\bf h}_X|^2 - |{\bf
       h}_Y|^2)\right] \,d\Omega, \label{SpQlin} \\
    U_{\rm sp} &=&
       {\lambda^2\over 4 \pi Z_0 } \int_\Omega
       \frac{|{\bf E}|^2}{2} \; \Re ({\bf h}_X \cdot {\bf
       h}_Y^*) \,d\Omega. \label{SpUlin}
\end{eqnarray}
Observing that
\begin{eqnarray}
   {\bf h}_{L}  &=& \frac{1}{\sqrt{2}}
                    \left({\bf h}_X - j {\bf h}_Y \right), \nonumber\\
   {\bf h}_{R}  &=& \frac{1}{\sqrt{2}}
                    \left({\bf h}_X + j {\bf h}_Y \right),
                    \label{Lin2Circ}
\end{eqnarray}
it is easy to recognize that
\begin{equation}
{\bf h}_R \cdot {\bf h}_L^* =
         \frac{1}{2}(|{\bf h}_X|^2 - |{\bf
          h}_Y|^2) + j\; \Re ({\bf h}_X \cdot {\bf
          h}_Y^*). \label{Pi1}
\end{equation}
Thus, the contamination on $Q$ and $U$ is a characteristic  of
the antenna system independent of 
the technique adopted to detect the two linear Stokes
parameters and a general contamination equation can be written
\begin{equation}
    Q_{\rm sp} + j\,U_{\rm sp} =
       {\lambda^2\over 4 \pi Z_0 } \int_\Omega
       {|{\bf E}|^2 \over 2}\; \Pi
       \;d\Omega ,
       \label{SpPi}
\end{equation}
where
\begin{eqnarray}
 \Pi &=& {\bf h}_R \cdot {\bf h}_L^* \nonumber \\
     &=& \frac{1}{2}(|{\bf h}_X|^2 - |{\bf
          h}_Y|^2) + j\; \Re ({\bf h}_X \cdot {\bf
          h}_Y^*) \label{Pi}
\end{eqnarray}
describes the degree of contamination produced by the
antenna itself and acts as a sort of {\it instrumental
polarization pattern} function.

Let (${\bf \hat x'},~{\bf \hat y'},~{\bf \hat z'}$) 
be the antenna reference frame, 
with ${\bf \hat z'}$ along the main beam and 
(${\bf \hat x'}$,~${\bf \hat y'}$) defining the antenna aperture  plane. 
The radiation patterns of the antenna
are conveniently described (also from experimental point of view)
according to the linear polarization basis 
(Ludwig's III definition,
Ludwig 1973)
\begin{eqnarray}
    {\hat {\bf p}} &=& \cos \phi' \, {\hat{\bf e}_{\theta'}} -
                       \sin \phi' \, {\hat{\bf e}_{\phi'}},
         \nonumber \\
    {\hat {\bf q}} &=& \sin \phi' \, {\hat{\bf e}_{\theta'}} +
                       \cos \phi' \, {\hat{\bf e}_{\phi'}},
   \label{pqeq}
\end{eqnarray}
where $\theta'$ and $\phi'$ are the polar and the azimuthal angles
of the antenna reference frame, while ${\hat{\bf e}_{\theta'}}$
and ${\hat{\bf e}_{\phi'}}$ are the tangential vectors of the
polar basis. In this basis
${\bf h}_X$ and ${\bf h}_Y$  can be written in terms of the
co-polar $(g)$ and cross-polar $(\chi)$ patterns
\begin{eqnarray}
   {\bf h}_X  &=& g_x \, {\hat {\bf p}} +
                            \chi_x \, {\hat {\bf q}}, \nonumber \\
   {\bf h}_Y      &=& \chi_y\, {\hat {\bf p}} +
                             g_y\, {\hat {\bf q}}.
   \label{hxhyeq}
\end{eqnarray}
Now, by substituting these expressions into equation (\ref{Pi}),
the instrumental polarization pattern $\Pi(\theta',\phi')$ writes
\begin{equation}
    \Pi(\theta', \phi')        =
       \Pi_Q(\theta', \phi')+j \,\,\Pi_U(\theta', \phi')
\end{equation}
with
\begin{eqnarray}
        \Pi_Q
        &=&     { |g_x|^2 + |\chi_x|^2
                 -|g_y|^2 - |\chi_y|^2
                 \over 2}, \label{fQeq} \\
        \Pi_U
        &=&     \Re\,(g_x \chi_y^* +
                     g_y \chi_x^*).
                  \label{fUeq}
\end{eqnarray}
Finally, by expressing the incoming radiation intensity in terms
of the brightness temperature $T_b(\theta', \phi')$
\begin{equation}
\frac{|{\bf E(\theta', \phi')}|^2}{Z_0} = \frac{2\,k\,T_b(\theta',
\phi')}{\lambda^2},
\end{equation}
where $k$ is the Boltzmann constant, the $Q-U$ contamination can
be written in terms of antenna temperature
\begin{eqnarray}
    Q_{\rm sp} + j\,U_{\rm sp} &=&
       {1 \over 4 \pi} \int_\Omega T_b(\theta', \phi') \,\, \Pi(\theta',\phi')
                 \,d\Omega'. \label{tspZeq}
\end{eqnarray}
Equation (\ref{tspZeq}) suggests to view  the contamination map of
$Q$ and $U$ as  the  convolution of the unpolarized radiation
$T_b(\theta', \phi')$ with the instrumental polarization pattern
$\Pi(\theta', \phi')$. 

\begin{figure}
 \resizebox{\hsize}{!}{\includegraphics[width=1.0\hsize]{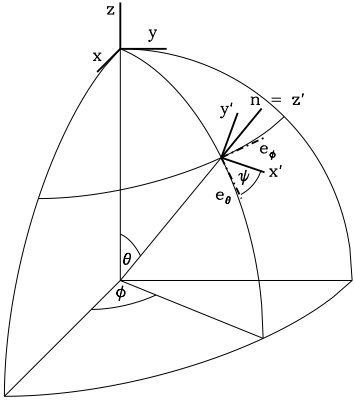}}
 \caption{The orientation of the instrument is defined by the Euler
angles ($\phi$,~$\theta$,~$\psi$) of the rotation to take the
instrument reference frame
($\bf{\hat x}'$,~$\bf{\hat y}'$,~$\bf{\hat z}'$)
onto the one fixed to the sky
($\bf{\hat x}$,~$\bf{\hat y}$,~$\bf{\hat z}$).
The contamination $Q_{\rm sp} + jU_{\rm sp}$, when the instrument
is pointing to the direction ${\bf \hat n}$, is obtained by
evaluating the brightness
temperature $T_b$ in the instrument reference frame. 
This is performed by the transformation taking the
sky reference frame onto that of the instrument.
}
 \label{refFrameFig}
\end{figure}

Equation~(\ref{tspZeq}) is valid in the 
antenna reference frame,
and provides the contamination when the instrument is pointing 
to the North Pole.
When the antenna  observes towards a generic direction $\bf{\hat
n} = (\theta, \phi)$ (see
Figure~\ref{refFrameFig}), a rotation has to be performed to account
for the instrument orientation relatively to the sky reference
frame, including a rotation by $\psi$ with respect to the polar
basis ($\bf{\hat e}_{\theta}$,~$\bf{\hat e}_{\phi}$).
%%%%%%%%%%%%%%%%%%%%%%%%%%%%%%%%%%%%%%%%%%%%%%%%%%%%%%%%%%%%%%%%%%%%%%%%
%%%%%%%%%%%%%%%%%%%%%%%%%%%%%%%%%%%%%%%%%%%%%%%%%%%%%%%%%%%%%%%%%%%%%%%%
In fact, referring to Figure~\ref{refFrameFig}, 
equation~(\ref{tspZeq}) requires the temperature field $T_b$
computed in the instrument reference frame 
(${\bf \hat x'},~{\bf \hat y'},~{\bf \hat z'} = {\bf \hat n}$).
Thus, given $T_b$ in the reference frame fixed to the sky
(${\bf \hat x},~{\bf \hat y},~{\bf \hat z}$),
one needs to take (${\bf \hat x},~{\bf \hat y},~{\bf \hat z}$)
onto (${\bf \hat x'},~{\bf \hat y'},~{\bf \hat z'}$)
by the three Euler's rotations: a first rotation around the 
$\bf{\hat z}$ axis by $\phi$, a second one
around the new $\bf{\hat y}$ axis by $\theta$ and a last rotation
around the new $\bf{\hat z}$ axis by $\psi$
(see also Challinor et al. 2000),
so that equation~(\ref{tspZeq}) transforms into
\begin{eqnarray}
        \left(Q_{\rm sp} + jU_{\rm sp}\right) (\theta,\phi)
        &=& {1 \over 4\pi} \int_\Omega
                 \left[R_{\bf{\hat z}}(\psi)
                       R_{\bf{\hat y}}(\theta)
                       R_{\bf{\hat z}}(\phi)
                       T_b\right](\theta',\phi')\nonumber\\
        & &      \Pi(\theta',\phi')
                 \,d\Omega', \nonumber\\
                 \label{tsp2eq}
\end{eqnarray}
where $\theta'$ and  $\phi'$ are the integration variables and
$R_{\bf{\hat r}}(\alpha)$ is the rotation operator around the
$\bf{\hat r}$ axis by an angle $\alpha$.
Alternatively, one has to express $\Pi$ in the sky
reference frame by the 
inverse of the transformation described above, so that
equation~(\ref{tspZeq}) also transforms in
\begin{eqnarray}
        \left(Q_{\rm sp} + jU_{\rm sp}\right) (\theta,\phi)
        &=& {1 \over 4\pi} \int_\Omega T_b(\theta',\phi')\nonumber\\
        & &      \left[R_{\bf{\hat z}}(-\phi)
                       R_{\bf{\hat y}}(-\theta)
                       R_{\bf{\hat z}}(-\psi)\;
                       \Pi\right] (\theta',\phi')
                 \,d\Omega'. \nonumber\\
                 \label{tspeq}
\end{eqnarray}
As anticipated, the result is the convolution between 
the unpolarized emission map and the
instrumental polarization pattern $\Pi$.

Equations~(\ref{tsp2eq})-(\ref{tspeq})
provide the intrumental polarization as measured
at the $Q$ and $U$ outputs of the instrument.
However, when building the polarized emission maps we must refer
$Q$ and $U$ to a standard reference frame. Here we adopt
the polar basis $\bf{\hat e}_{\theta}$ and
$\bf{\hat e}_{\phi}$ along meridians and parallels, 
respectively{\footnote{As a matter of fact,
the standard definition of polarization angles refers to 
$-\bf{\hat e}_{\phi}$ as $\bf{\hat y}$ axis (Berkuijsen 1975),
differing from that adopted here by
an axis reflection. In the reflected (Berkuijsen) 
reference frame the contamination equations are valid 
for $Q-jU$}} 
(see Figure~\ref{refFrameFig}). 
A counter-rotation
by $\psi$ of $Q+jU$ is thus required to evaluate the contamination
$Q_{\rm st} + jU_{\rm st}$ in the final maps
\begin{eqnarray}
        \left(Q_{\rm st} + jU_{\rm st}\right) (\theta, \phi)
        &=&  (Q_{\rm sp} + jU_{\rm sp})\, e^{j2\psi}\nonumber\\
        &=& {1 \over 4\pi} \int_\Omega
                 \left[R_{\bf{\hat z}}(\psi)
                       R_{\bf{\hat y}}(\theta)
                       R_{\bf{\hat z}}(\phi)
                       T_b\right](\theta', \phi')\nonumber\\
        & &      \Pi(\theta', \phi')\, e^{j2\psi}
                 \,d\Omega' \nonumber\\
        &=& {1 \over 4\pi} \int_\Omega
                 \left[R_{\bf{\hat y}}(\theta)
                       R_{\bf{\hat z}}(\phi)
                       T_b\right](\theta'', \phi'')\nonumber\\
        & &      \Pi(\theta'', \phi''-\psi)\, e^{j2\psi}
                 \,d\Omega'', \nonumber\\
                 \label{tspSteq}
\end{eqnarray}
where $(\theta', \phi')$ and $(\theta'', \phi'')$ are the coordinates in the
instrument and standard frames, respectively, related by
\begin{eqnarray}
   \theta'' &=& \theta', \nonumber\\
   \phi''   &=& \phi' + \psi.
\end{eqnarray}
Equations~(\ref{tspZeq}-\ref{tspSteq}) provide a complete computation
of the instrumental polarization generated by the antenna and extend
equation~(A.6) in Carretti~et~al.~(2001),
where only the effect on $U$ with the antenna pointing towards the North Pole 
was computed.
%%%%%%%%%%%%%%%%%%%%%%%%%%%%%%%%%%%%%%%%%%%%%%%%%%%%%%%%%%%%%%%%%%%%%%%%
\begin{figure*}
 \centering
{\includegraphics[width=0.45\hsize]{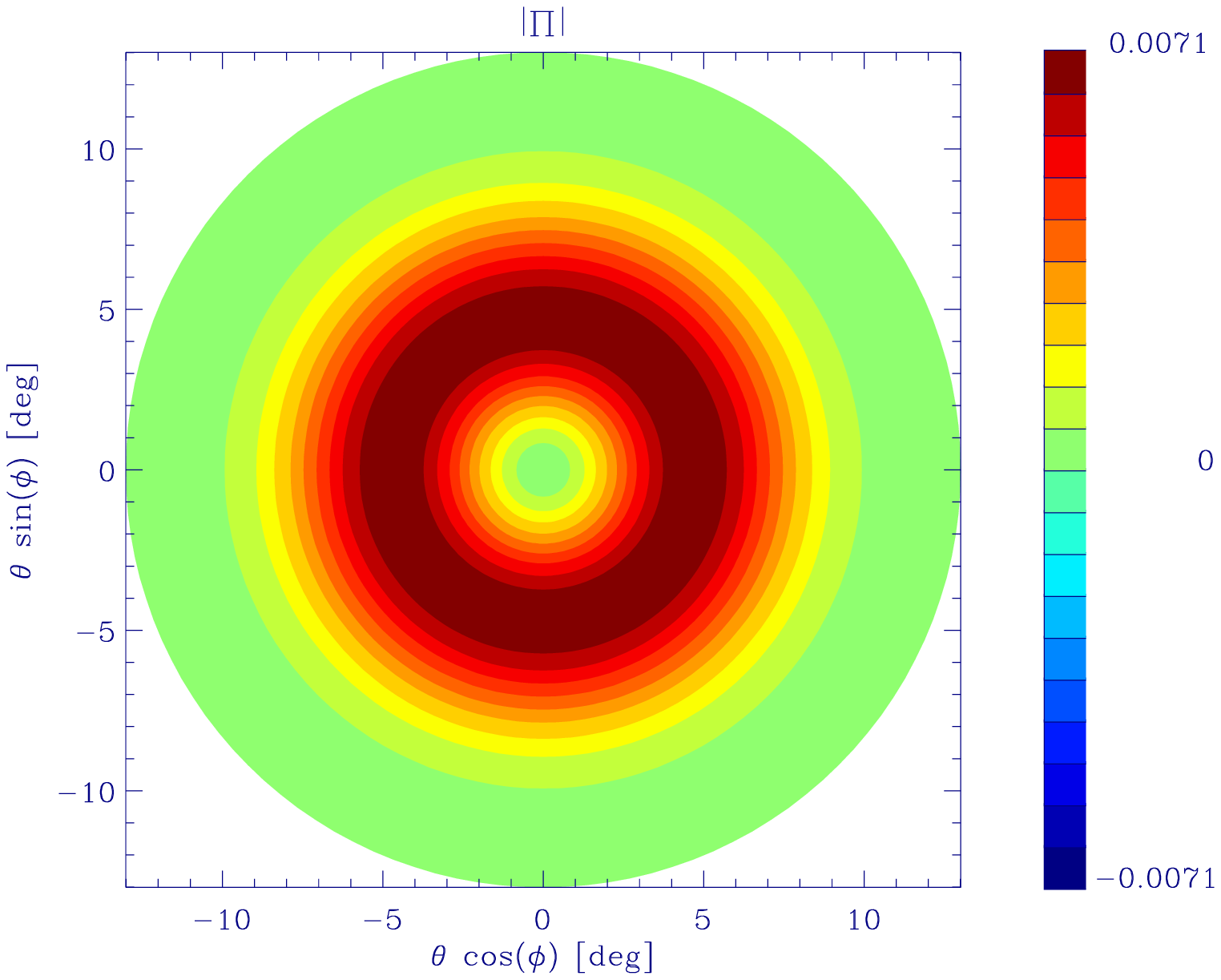}}
{\includegraphics[width=0.45\hsize]{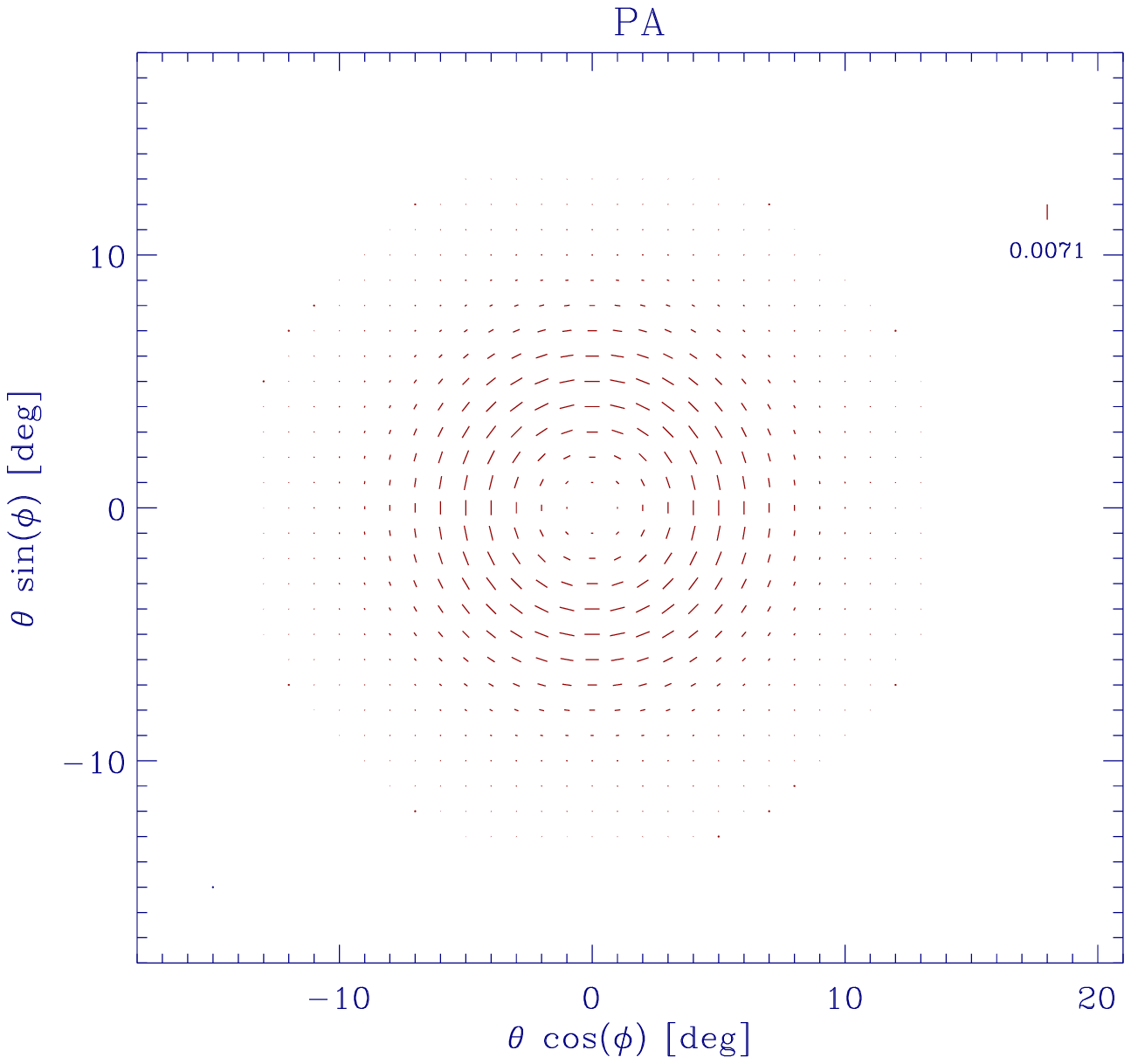}}\\
{\includegraphics[width=0.45\hsize]{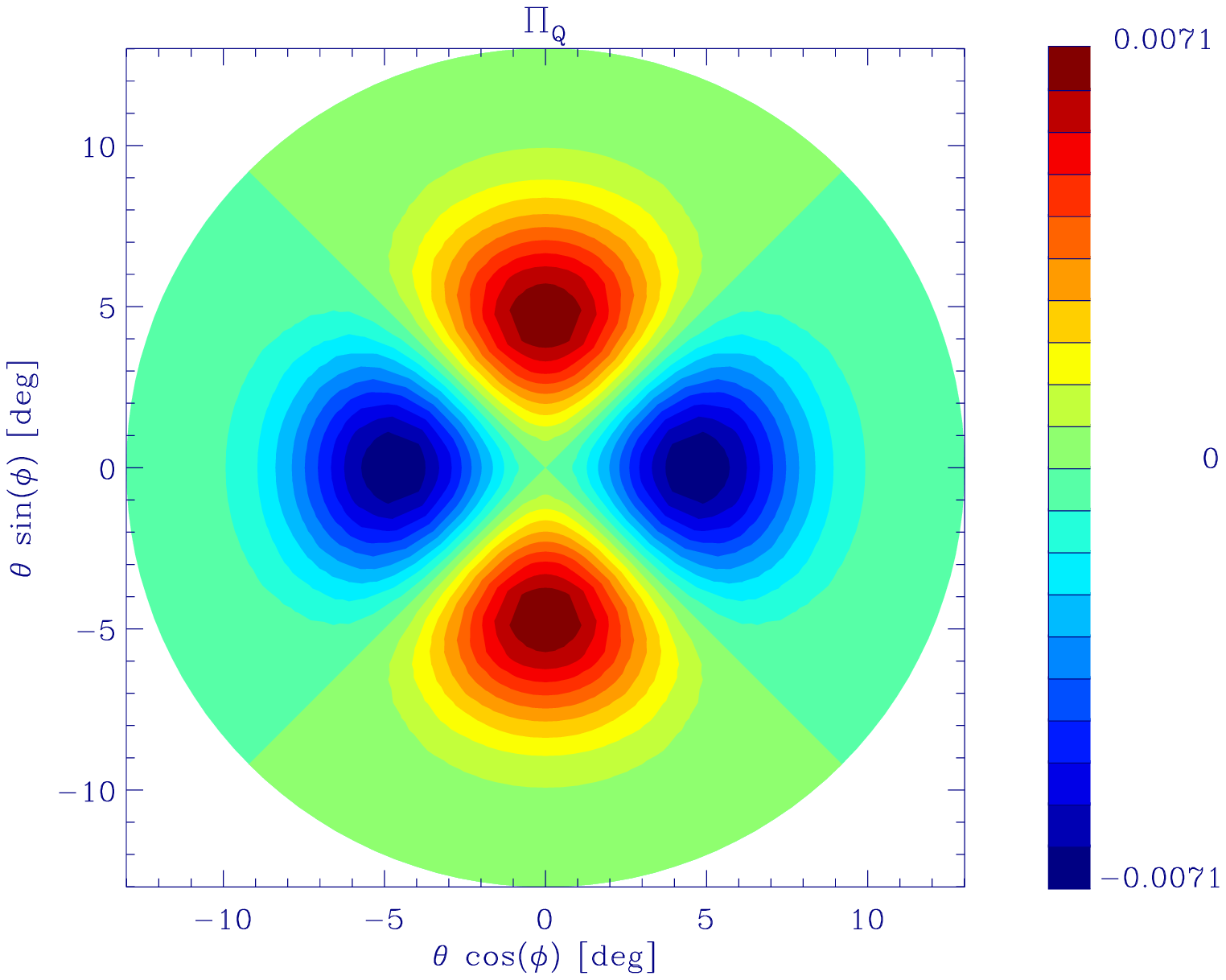}}
{\includegraphics[width=0.45\hsize]{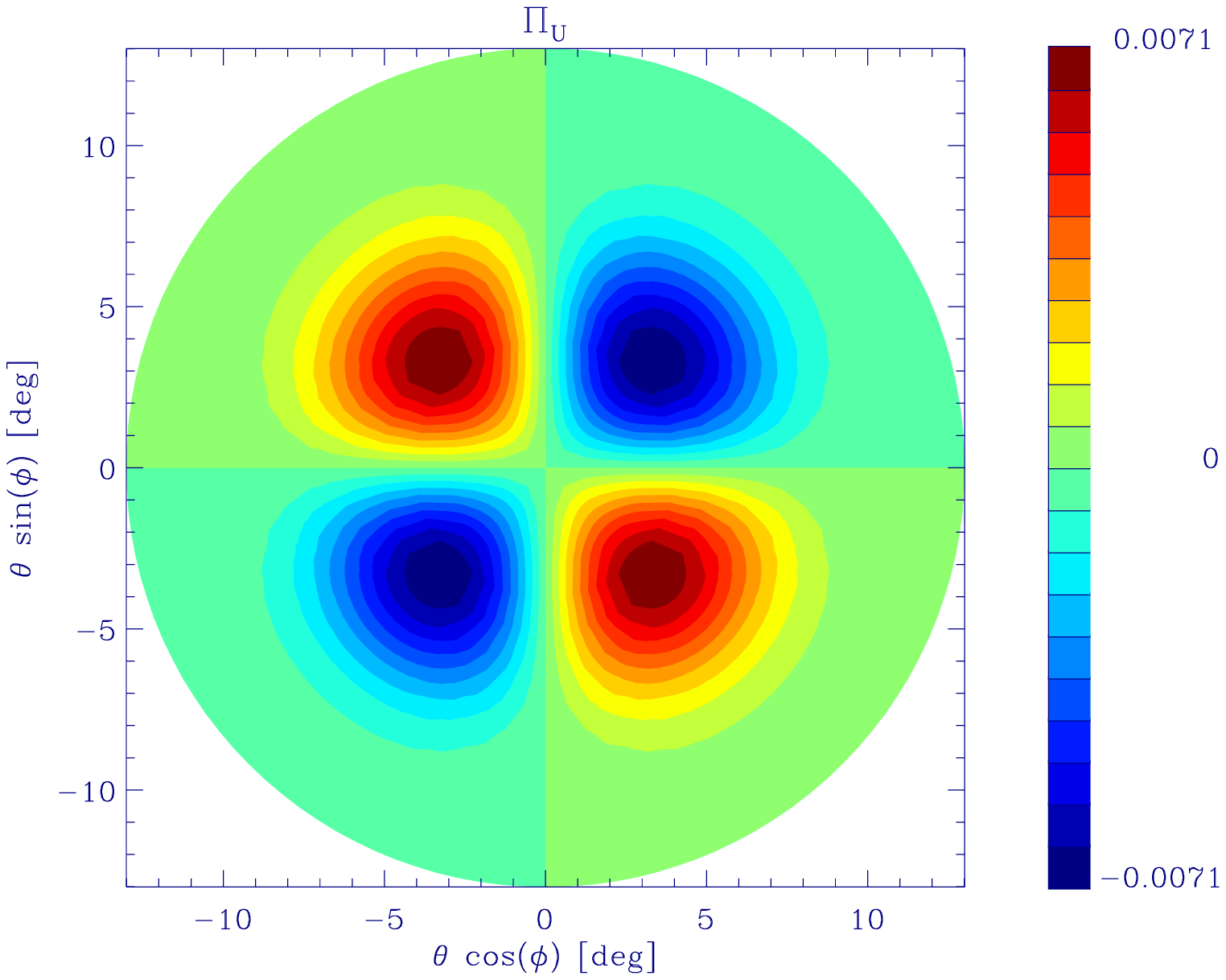}}
 \caption{Instrumental polarization beams
 normalized to the co-polar maximum
 for the 90 GHz feed horn
 of the SPOrt experiment. The $|\Pi|$ map (top left) shows
 the axial symmetry of the beam, while the polarization angle map
 (top right: the vector
 length is proportional to the intensity) presents a radial pattern.
 In the case of SPOrt the polarization angles are tangential
 to the radial direction. The instrumental polarization beams $\Pi_Q$ (bottom left) and
 $\Pi_U$ (bottom right) have quadrilobe patterns that are
 symmetric with respect to the
 main axes; the change of sign from quadrant to quadrant makes the instrumental
 contamination in both $Q$ and $U$ only sensitive to the anisotropy pattern
 of the unpolarized radiation (see text for details).}
 \label{SPOrtbeamMap}
\end{figure*}
%%%%%%%%%%%%%%%%%%%%%%%%%%%%%%%%%%%%%%%%%%%%%%%%%%%%%%%%%%%%%%%%%%%%%%%%

\section{Axisymmetric Case}\label{axSymmSec}

Axial symmetry is an important special case representing several common
optics solutions. Circular dual--po\-la\-ri\-za\-tion feed horns and
on--axis mirror configurations like Casse\-grain are some examples.
The axial symmetry leads to some
simplifications in the instrumental polarization
equations with interesting effects on the $E$
and $B$-mode of the spurious polarization.
In details, by considering the symmetric property 
of the radiation aperture, one
co-polar pattern $g$ and one cross-polar
$\chi$ are needed and the following relations hold  (Rahmat-Samii 1993)
\begin{eqnarray}
  g_{x} (\theta,\,\phi) &=& g(\theta,\,\phi) \nonumber\\
  g_{y}     (\theta,\,\phi) &=& g(\theta,\,\phi - \pi/2) \nonumber\\
  \chi_{x} (\theta,\,\phi)  &=& \chi (\theta,\,\phi) \nonumber\\
  \chi_{y}  (\theta,\,\phi) &=& -\chi (\theta,\,\phi -\pi/2) \label{gchieq}
\end{eqnarray}
together with the periodic and parity properties
\begin{eqnarray}
  g(\theta,\,\phi) &=& g(\theta,\,\phi+\pi) \nonumber\\
  g(\theta,\,\phi) &=& g(\theta,\,-\phi) \nonumber\\
  \chi (\theta,\,\phi) &=& \chi (\theta,\,\phi+\pi) \nonumber\\
  \chi (\theta,\,\phi) &=& -\chi (\theta,\,-\phi) \label{parityeq}
\end{eqnarray}
where $\theta$ is the angular distance from the axis and $\phi$ the azimuthal
angle with respect to the instrument 
Cartesian reference frame centred on the main
axis.

Considering these properties, equation~(\ref{tspZeq}) writes
\begin{eqnarray}
        Q_{\rm sp} + j\,\,U_{\rm sp}
        &=& {1 \over 4\pi} \int_0^{\pi} \sin\theta\,d\theta
                               \int_0^{\pi\over2}d\phi \cdot \nonumber\\
        & &  \left[ T_b(\theta, \phi)- T_b(\theta, \phi+{\pi\over 2})\; + \right.\nonumber\\
        & &  \left. T_b(\theta, \phi+\pi)- T_b(\theta,\phi+{3\over2}\pi)\right]\cdot\nonumber\\
        & &  \left[\Pi_Q(\theta, \phi) + j\,\Pi_U(\theta, \phi)\right]
                  \label{symmtspeq}
\end{eqnarray}
where the integration is only performed  over the first quadrant.
The contribution of a constant $T_b$ term is null and
the instrumental polarization in both $Q$ and $U$
only depends on the anisotropy of the unpolarized emission.
This result extends
equation~(18) in Carretti~et~al.~(2001) to $Q$.

Further properties of the axisymmetric case lead to a better
understanding of the
nature of this contamination and the instrumental polarization beam
$\Pi=\Pi_Q+j\Pi_U$. In fact, $g$ and $\chi$ have the simple expressions
\begin{eqnarray}
  g(\theta,\,\phi) &=& g(\theta,\,\phi=0)\,\cos^2(\phi) +
                       g(\theta,\,\phi={\pi\over 2})\,\sin^2(\phi)\nonumber\\
                   &=& g_0(\theta)\,\cos^2(\phi) +
                       g_{\pi/2}(\theta)\,\sin^2(\phi) \\
                       \nonumber\\
  \chi (\theta,\,\phi) &=&  {g_0(\theta) - g_{\pi/2}(\theta)\over 2} \sin(2\phi)
          \label{gogpieq}
\end{eqnarray}
so that the instrumental polarization beams of $Q$ and $U$
result in
\begin{eqnarray}
  \Pi_Q (\theta,\,\phi) &=&  {|g_0(\theta)|^2 - |g_{\pi/2}(\theta)|^2 \over 2}
                           \;\cos(2\phi) \label{symmfQeq}\\
  \Pi_U (\theta,\,\phi) &=&  {|g_0(\theta)|^2 - |g_{\pi/2}(\theta)|^2 \over 2}
                           \;\sin(2\phi)  \label{symmfUeq}
\end{eqnarray}
and the contamination equation in
\begin{equation}
        Q_{\rm sp} + j\,\,U_{\rm sp}
        = {1 \over 4\pi} \int_\Omega
             T_b(\theta, \phi)\
            {|g_0|^2 - |g_{\pi/2}|^2 \over 2}\,e^{j2\phi}
                 \,d\Omega. \label{symmtsp2eq}
\end{equation}
The most relevant property of the instrumental polarization beam
\begin{equation}
        \Pi = \Pi_Q + j\Pi_U
        = {|g_0(\theta)|^2 - |g_{\pi/2}(\theta)|^2 \over 2}\,e^{j\,2\phi}
        \label{symmfeq}
\end{equation}
is the axial symmetry, with intensity
only depending on the angular distance $\theta$ from the axis 
\begin{equation}
        |\Pi|
        = \left|{|g_0(\theta)|^2 - |g_{\pi/2}(\theta)|^2 \over 2}\right|
        \label{symmPfeq}
\end{equation}
and {\it polarization angle}
with radial pattern with respect to the beam axis
\begin{eqnarray}
        \alpha
        &=& 0.5 \arctan \;{U_{\rm sp} \over Q_{\rm sp}} \nonumber\\
        &=& \left\{
            \begin{array}{ll}
                 \phi & {\rm for\,|g_0|^2 > |g_{\pi/2}|^2}\\
                 \phi+90^\circ & {\rm otherwise}.\\
            \end{array}
        \right.
\end{eqnarray}
The angle $\alpha$ is directed either along the radial or the
tangential direction depending on
$|g_0|^2$ being larger or smaller than $|g_{\pi/2}|^2$.
Moreover, the instrumental polarization is given by the difference between the
co-polar cuts along the two main axes and it is thus related to
asymmetries like differences in the FWHMs along the two axes.
%%%%%%%%%%%%%%%%%%%%%%%%%%%%%%%%%%%%%%%%%%%%%%%%%%%%%%%%%%%%%%%%%%%%%%%%
 \begin{figure}
 \resizebox{\hsize}{!}{\includegraphics[width = 1.0\hsize, angle = 0]
     {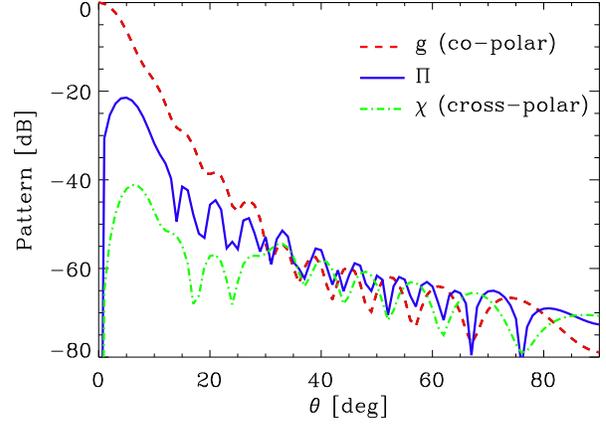}}
\caption{Amplitude of the instrumental
polarization pattern $\Pi$ for the 90 GHz feed horn
of the SPOrt experiment along a radial cut. The pattern is normalized to the
maximum of the main beam.
The co-polar and the cross-polar beams along the 45$^\circ$ cut are also
shown for
comparison.}
\label{SPOrtprof}
\end{figure}
%%%%%%%%%%%%%%%%%%%%%%%%%%%%%%%%%%%%%%%%%%%%%%%%%%%%%%%%%%%%%%%%%%%%%%%%

As an example, Figure~\ref{SPOrtbeamMap} shows the case of the 90~GHz
horn of SPOrt (Carretti et al. 2003).
SPOrt is an experiment devoted to measure the Cosmic Microwave
Background Polarization on large angular scales.
It uses very simple antennae like circular
feed horns with  FWHM~=~$7^\circ$ resolution,
intrisically axisymmetric.
The axial symmetry
of the instrumental polarized beam and the radial pattern of the
polarization angles are clearly visible in the $|\Pi|$ and polarization
angle maps.
The $Q$ and $U$ responses present quadrilobe patterns and
the change of sign with the
quadrants makes the contamination in both $Q$ and $U$ only sensitive to the
anisotropy of the unpolarized radiation
(equation~(\ref{symmtspeq})).
It is to be noted that $Q$ has the
quadrilobe pattern along the main axes, while $U$ along the $45^\circ$
directions.

Figure~\ref{SPOrtprof} shows the section of the
quantity $|\Pi|$ along one radial cut
normalized to the co-polar beam maximum
and provides the polarized contamination level as a function of the axial
distance. The maximum is at about FWHM/2 from the antenna axis and
the area where its action is effective has a diameter of about
$2\times{\rm FWHM}$.

Finally, the representation in terms of amplitude and
polarization angles of the contamination
yields that the contaminations
in both $Q$ and $U$ have the same nature
and are simply the two components of a unique 2-D quantity.

\section{Effects on $E$ and $B$-mode Power Spectra
         for the Axisymmetric Case}\label{EBSec}

The $\Pi$ pattern is the response of the system to an
unpolarized point source,
the contamination being given by its convolution with $T_b$.
In the axisymmetric case, $\Pi$ is axisymmetric in itself
with polarization angle either parallel or
perpendicular to the radial direction, suggesting that
a pure $E$-mode is present with no contribution to $B$.
To investigate the polarization power spectra of $\Pi$ we consider
the simple case of an antenna pointing to the North Pole.

To compute the $E$- and $B$-mode spectra we have
to write the $\Pi$ pattern in the polar basis where 
the spherical harmonics are referred to.
The transformation is performed by
parallel transporting the polarization vector along the great circle through
the poles (Ludwig 1973)
\begin{equation}
        \Pi^{\rm st}_Q + j \Pi^{\rm st}_U = (\Pi_Q + j\Pi_U) \,e^{-j2\phi}.
\end{equation}
The field $\Pi$ in the standard reference frame results thus in
a $Q$ component depending only on $\theta$ and a null $U$ term
\begin{eqnarray}
        \Pi^{\rm st}_Q &=& \Pi_Q(\theta,0) = {|g_0(\theta)|^2 - |g_{\pi/2}(\theta)|^2
\over 2},
                 \nonumber\\
        \Pi^{\rm st}_U &=& 0.
\end{eqnarray}
From Zaldarriaga \& Seljak (1997), for a field depending just on
$\theta$ the only non-null 2-spin harmonic coefficients are
$a_{\pm 2,\ell 0}$, which, in case of $U=0$, are real. Considering
the relation $a_{2,\ell m} = a_{-2,\ell -m}^*$, the field $\Pi$ has
$a_{2,\ell m} = a_{-2,\ell m}$ leading to the $E-$ and $B$--mode $\Pi$ coefficients
\begin{eqnarray}
        a_{E,\ell m}^\Pi &=& -a_{2,\ell 0}^\Pi \,\delta_{m\,0}
                \label{faElmeq}\nonumber\\
        a_{B,\ell m}^\Pi &=& 0
\end{eqnarray}
and confirming that $\Pi$ has no $B$-mode in the axisymmetric case.
Finally, the power spectra are
\begin{eqnarray}
        C_{E\ell}^\Pi &=& {1\over 2\ell+1} |a_{2,\ell0}^\Pi|^2,\nonumber\\
        C_{B\ell}^\Pi &=& 0. \label{fspeceq}
\end{eqnarray}
%%%%%%%%%%%%%%%%%%%%%%%%%%%%%%%%%%%%%%%%%%%%%%%%%%%%%%%%%%%%%%%%%%%%%%%%
\begin{figure}
 \resizebox{\hsize}{!}
 {\includegraphics[width = 1.0\hsize, angle = 0]{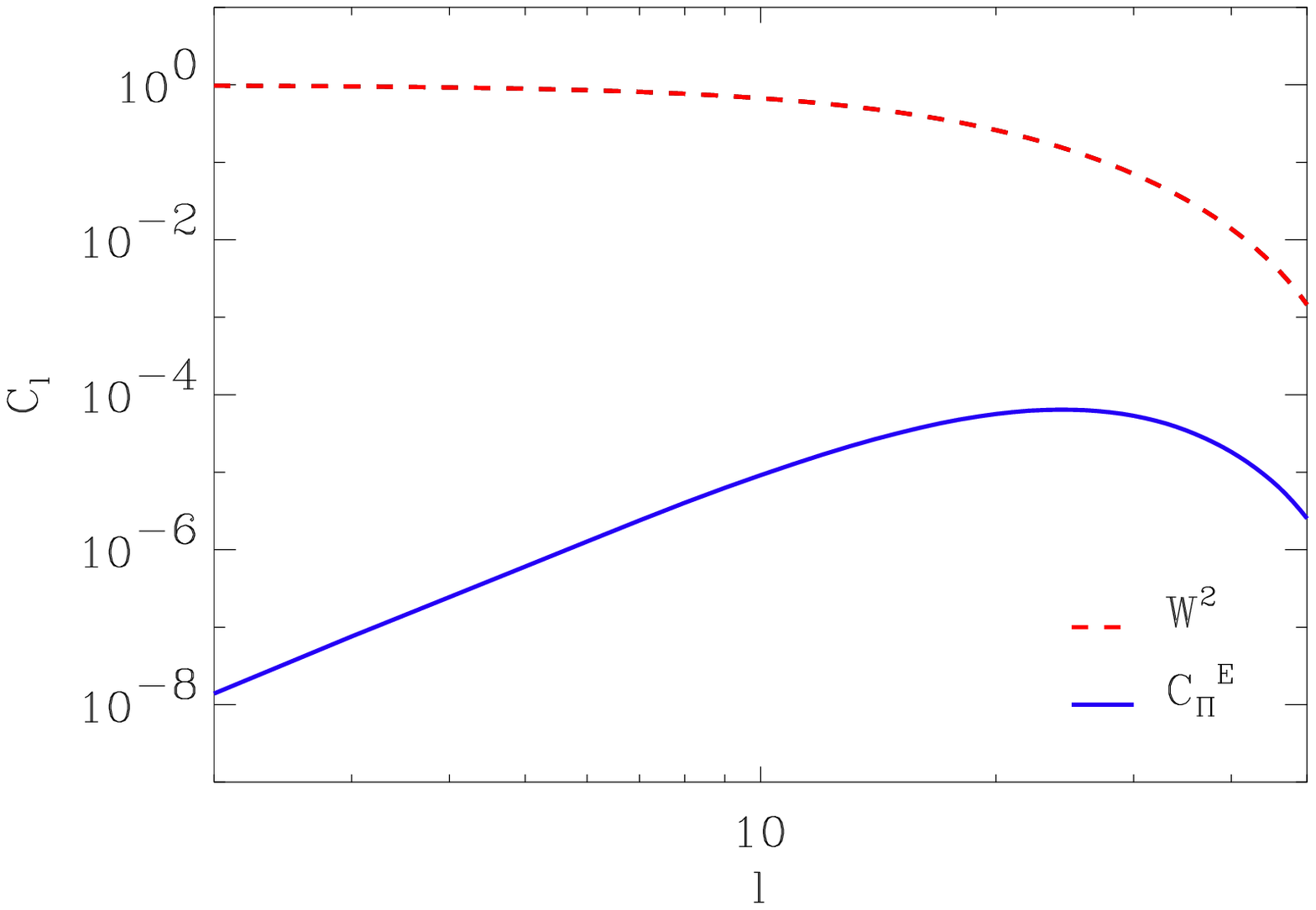}}
 {\includegraphics[width = 1.0\hsize, angle = 0]{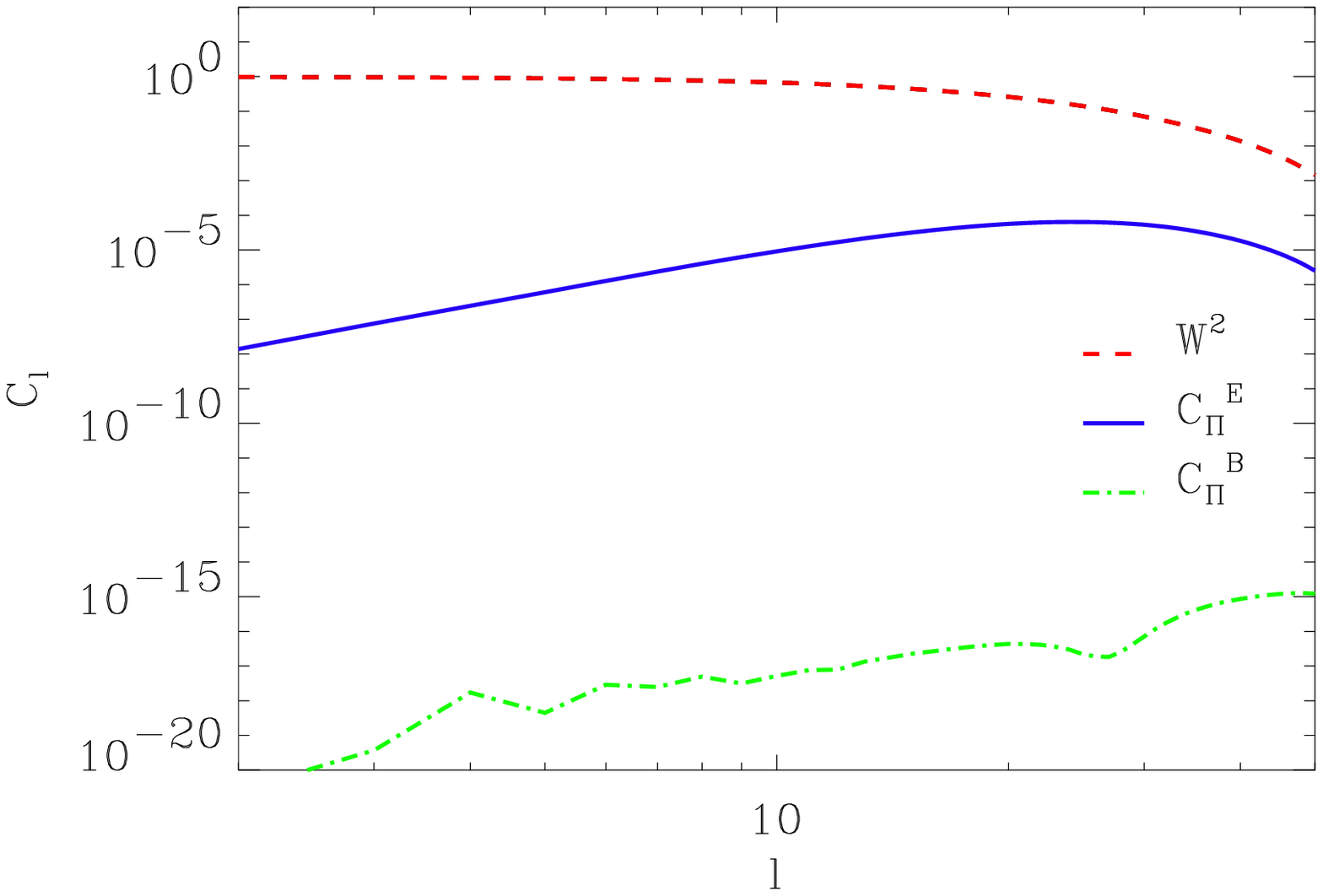}}
\caption{Top: $E$-mode power spectrum $C_E^\Pi$ of the instrumental
polarization beam $\Pi$  for the SPOrt 90 GHz horn
normalized to the spectrum of the window function $W_\ell^2$.
The $E$-mode peaks nearly at the multipole typical
of the FWHM ($\ell \sim 25$) leaving the larger scales significantly
cleaner. Note that at $\ell=2$ the $E$ spectrum is about 4 orders of
magnitude lower than at the peak and 8 orders of magnitude lower than the
window function of the main beam. Bottom: the same plot with
a larger range so to include the $B$-mode as well. Its level
is much lower than the $E$ one and its presence is likely due to numerical errors. 
In any case, it
is 12-13 orders of magnitude lower than the $E$-mode.}
\label{SPOrtBeamSpec}
\end{figure}
%%%%%%%%%%%%%%%%%%%%%%%%%%%%%%%%%%%%%%%%%%%%%%%%%%%%%%%%%%%%%%%%%%%%%%%%

Figure~\ref{SPOrtBeamSpec} shows the spectra computed for the
case of SPOrt horns.
Beyond the very low level of the computed $B$-mode
(non-zero value is likely due to numerical errors) an important property
of the $E$-mode appears: its spectrum
peaks at high $\ell$, approximately on the FWHM scale, and rapidly decreases
at smaller $\ell$ leaving the largest scales significantly cleaner.
This is a very important feature for instruments looking for
CMBP on large scales,
the relevant information being at $\ell < 10$ 
(e.g. see Zaldarriaga, Spergel \& Seljak 1997).

Besides the properties of $\Pi$,
to understand the impact of instrumental polarization on CMBP
experiments we have to
evaluate the instrumental polarization map
generated by a diffuse unpolarized emission.

In general, given a temperature map, equation~(\ref{tspSteq})
states that the contamination field
in not uniquely defined,
depending on the rotation of the instrument with respect to
the standard frame and, thus,
on the scanning strategy.

This ambiguity is lost in the axisymmetric case, for which
 equation~(\ref{symmfeq}) holds and
 equation~(\ref{tspSteq}) writes
\begin{eqnarray}
        Q_{\rm st} + jU_{\rm st}
        &=& {1 \over 4\pi} \int_\Omega
                 \left[R_{\bf{\hat y}}(\theta)
                       R_{\bf{\hat z}}(\phi)
                       T_b\right](\theta'', \phi'')\nonumber\\
        & &      \Pi(\theta'',0)\, e^{j2\phi''}
                 \,d\Omega'', \nonumber\\
                 \label{symmtspSteq}
\end{eqnarray}
which is totally independent of the instrument rotation $\psi$
and corresponds to the contamination for an antenna
aligned with the standard frame ($\psi=0$).
Axisymmetric antennae have thus the interesting
feature of generating instrumental polarization maps that are
independent of the scanning strategy.
%%%%%%%%%%%%%%%%%%%%%%%%%%%%%%%%%%%%%%%%%%%%%%%%%%%%%%%%%%%%%%%%%%%%%%%%
\begin{figure}
  \resizebox{\hsize}{!}{\includegraphics[width=1.0\hsize]{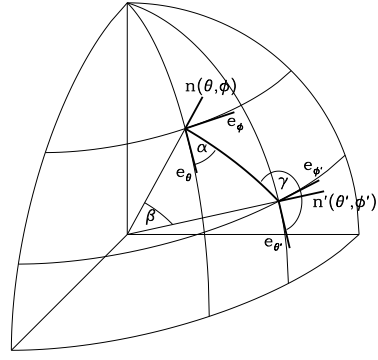}}
 \caption{The position of
$\bf{\hat n} =(\theta, \phi)$ with respect to
 $\bf{\hat n'} =(\theta', \phi')$ is
defined by three angles: $\beta$, the angular separation
between the two versors; $\alpha$, the angle to take
$\bf{\hat e}_{\theta}({\bf{\hat n}})$ in right-handed sense
onto the great circle connecting $\bf{\hat n}$ and $\bf{\hat n'}$;
$\gamma$, the same as $\alpha$ but referred to
$\bf{\hat e}_{\theta}({\bf{\hat n'}})$.
}
 \label{refABGFig}
\end{figure}
%%%%%%%%%%%%%%%%%%%%%%%%%%%%%%%%%%%%%%%%%%%%%%%%%%%%%%%%%%%%%%%%%%%%%%%%

From equation~(\ref{symmfeq}) and
considering $\psi=0$, the contamination
equation~(\ref{tspeq}) writes
\begin{equation}
        \left(Q_{\rm sp} + jU_{\rm sp} \right) ({\bf{\hat n}})
        = {1 \over 4\pi} \int_\Omega T_b({\bf{\hat n'}}) \Pi(\beta, 0) e^{j2\alpha}
                 \,d\Omega' \nonumber\\
                 \label{tspalphaeq}
\end{equation}
with $\beta$ the angle between the directions 
${\bf{\hat n}}=(\theta,\phi)$ and 
${\bf{\hat n'}}=(\theta',\phi')$
and $\alpha$ the angle to rotate in the right-handed sense
the versor ${\bf{\hat e}_{\theta}}({\bf{\hat n}})$ onto the great
circle connecting ${\bf{\hat n}}$ and ${\bf{\hat n'}}$ (see
Figure~\ref{refABGFig}).
%%%%%%%%%%%%%%%%%%%%%%%%%%%%%%%%%%%%%%%%%%%%%%%%%%%%%%%%%%%%%%%%%%%%%%%%
\begin{figure*}
 \centering
{\includegraphics[width = 0.35\hsize, angle = 90]{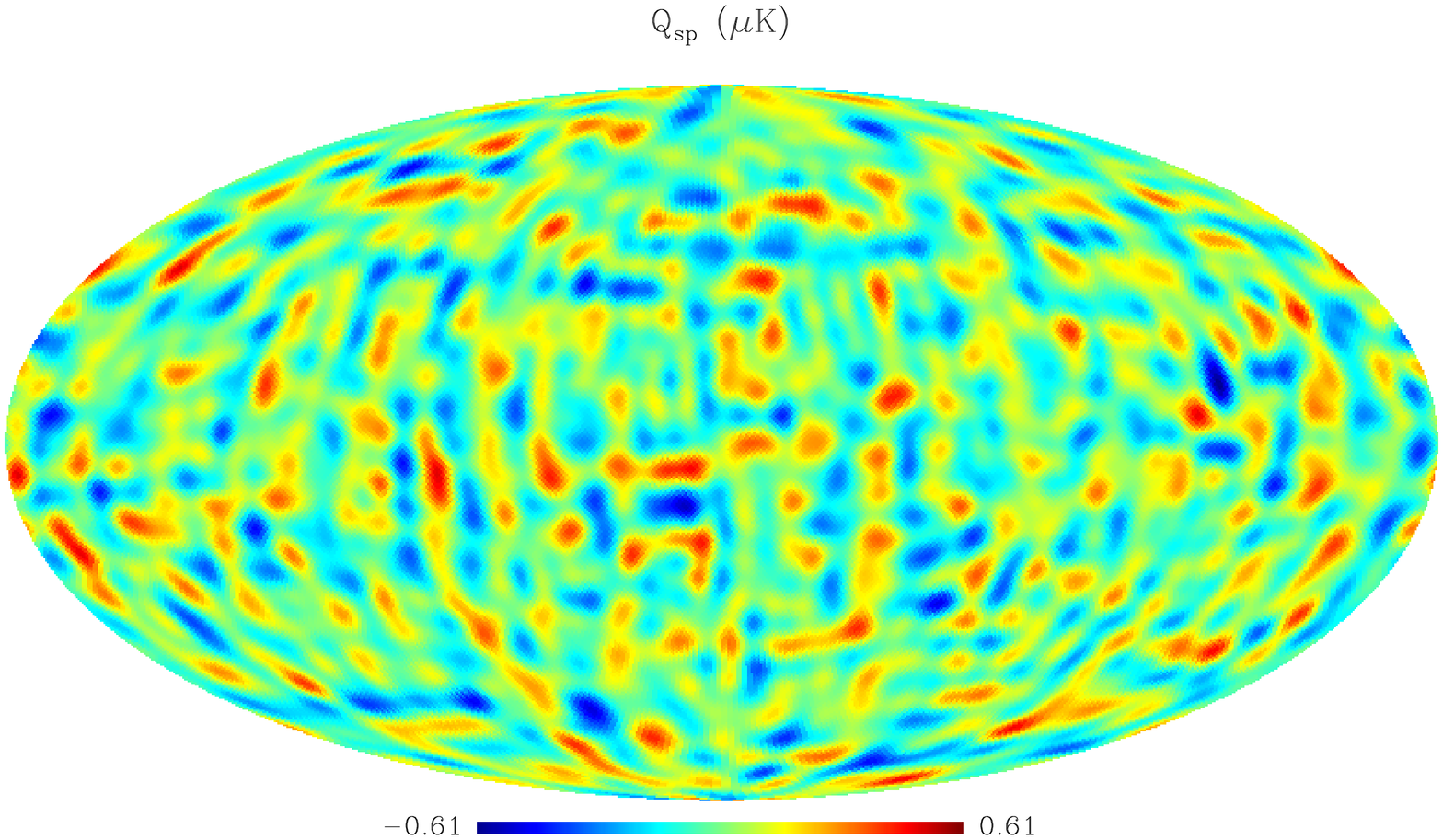}}
{\includegraphics[width = 0.35\hsize, angle = 90]{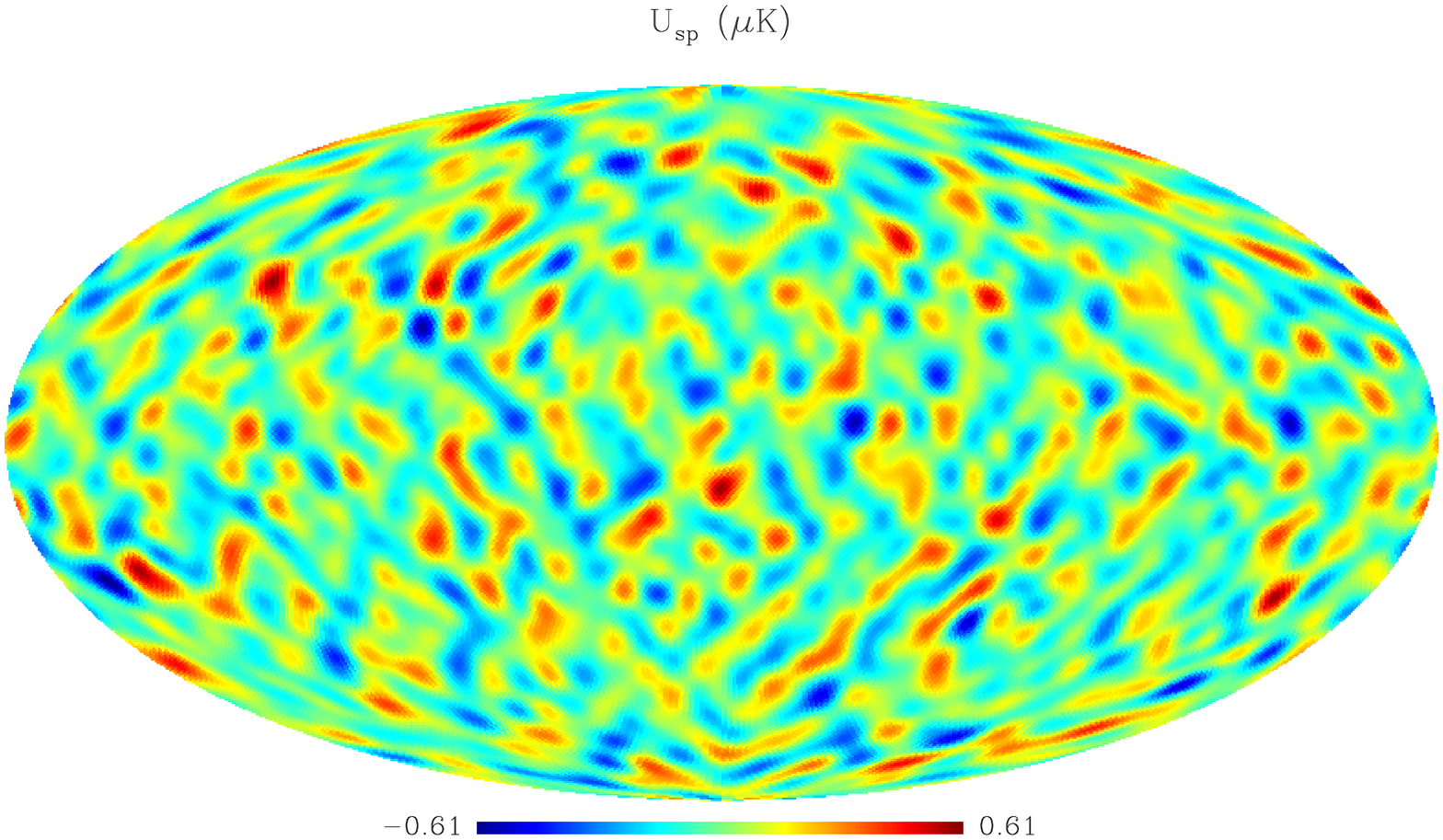}}
{\includegraphics[width = 0.35\hsize, angle = 90]{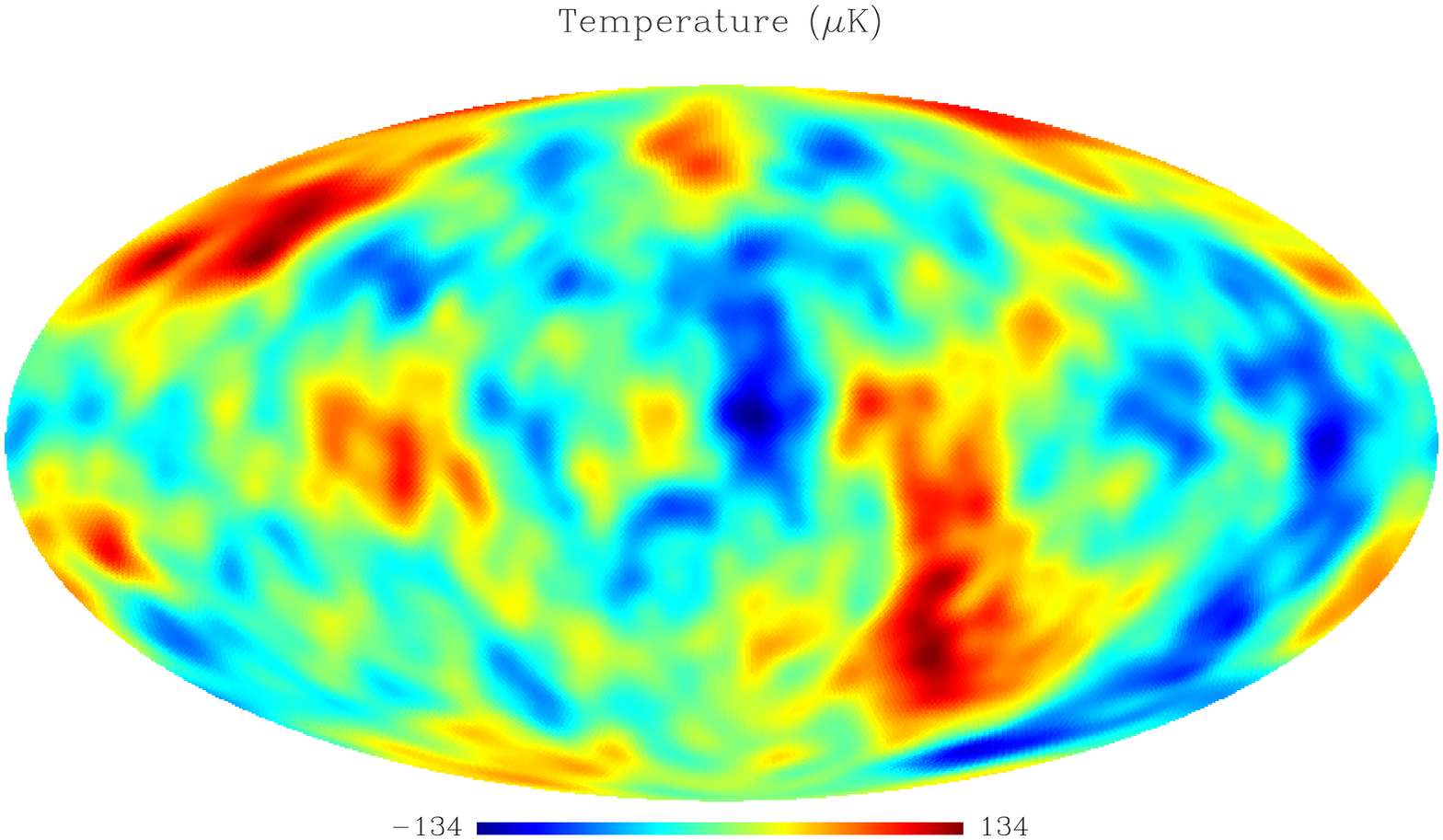}}
\caption{Contamination maps for $Q$(top) and $U$(mid) generated
by the convolution between the $\Pi$ beam of the SPOrt horn and
the Temperature map of a $\Lambda$CDM model (concordance model
as from WMAP first year results).
The Temperature map, smeared by the FWHM~$\sim 7^\circ$
beam, is also shown (bottom).}
\label{contMapFig}
\end{figure*}
%%%%%%%%%%%%%%%%%%%%%%%%%%%%%%%%%%%%%%%%%%%%%%%%%%%%%%%%%%%%%%%%%%%%%%%%

Making use of the relation (Ng \& Liu 1999)
\begin{eqnarray}
Y_{s,\ell m}(\theta, \phi) = \sqrt{4\pi \over 2\ell+1} \sum_{m'} Y_{s,\ell m'}(\beta, \gamma)
                         Y_{-m',\ell m}(\theta', \phi')\,e^{js\alpha}
                         \nonumber\\
\end{eqnarray}
with  $\gamma$ equivalent to $\alpha$ but for the direction ${\bf{\hat n'}}$,
the $a_{2,\ell m}^{T\otimes \Pi}$ coefficient of the contamination map results in
\begin{eqnarray}
        a_{2,\ell m}^{T\otimes \Pi}
        &=& \int d\Omega\,
                   (Q_{\rm sp} + jU_{\rm sp})({\bf{\hat n}})\;
                   Y_{2,\ell m}^*({\bf{\hat n}}) \nonumber\\
        &=& {1 \over \sqrt{4\pi \, \left(2\ell+1\right)}}
             \sum_{m'}\int d\Omega'\,T_b({\bf{\hat n'}})
             Y_{-m',\ell m}^*({\bf{\hat n'}})\cdot \nonumber\\
        & & \int d\Omega_{\beta\gamma}\, Y_{2,\ell m'}^*(\beta,\gamma) 
            \Pi(\beta,0) \nonumber\\
        &=& {1 \over \sqrt{4\pi \, \left(2\ell+1\right)}}
             \sum_{m'}\int d\Omega'\,T_b({\bf{\hat n'}})
             Y_{-m',\ell m}^*({\bf{\hat n'}})\; a_{2,\ell m'}^\Pi \nonumber\\
\end{eqnarray}
where $d\Omega_{\beta\gamma}$ denotes the integration
on the sphere centred on ${\bf{\hat n'}}$. Since the last integral is the 
Since $a_{2,\ell m}^\Pi = a_{2,\ell 0}^\Pi \delta_{m0}$, it is straighforward
\begin{eqnarray}
        a_{2,\ell m}^{T\otimes \Pi}
        &=& {a_{2,\ell 0}^\Pi \over \sqrt{4\pi \, \left(2\ell+1\right)}}
             \int d\Omega'\,T_b({\bf{\hat n'}})
             Y_{\ell m}^*({\bf{\hat n'}}) \nonumber\\
        &=& {1 \over \sqrt{4\pi \, \left(2\ell+1\right)}}\,\,
            {a_{2,\ell 0}^\Pi \,a_{\ell m}^T}.
\end{eqnarray}
Similarly,
\begin{eqnarray}
        a_{-2,\ell m}^{T\otimes \Pi}
        &=& {1 \over \sqrt{4\pi \, \left(2\ell+1\right)}}\,\,
            {a_{-2,\ell 0}^\Pi \,a_{\ell m}^T},
\end{eqnarray}
leading to the power spectra of the contamination map
\begin{eqnarray}
     C_{E\ell}^{T\otimes \Pi} &=&  {1 \over 4\pi} \, C_{E\ell}^\Pi \, C_{T\ell},
      \nonumber\\
     C_{B\ell}^{T\otimes \Pi} &=& 0. \label{cecbeq}
\end{eqnarray}
Thus, there is no $B$-mode component
in the contamination map, and
the $E$ power spectrum is the product between
the spectra of the Temperature map $C_{T\ell}$ and
the beam $\Pi$, making valid a sort
of convolution theorem.
Note that this result can be applied to all antenna systems 
with axisymmetric $\Pi$ pattern, even if, in general, this cannot be stated
for off-axis optics.

To evaluate the effects of an axisymmetric optics
on a CMBP experiment we take
the example of the SPOrt instrument.
The contamination maps $Q_{\rm sp}$ and $U_{\rm sp}$
are computed
assuming a CMB anisotropy map
compatible with the concordance model
of the WMAP first-year data
(Bennett et al. 2003, Spergel et al. 2003).
We convolve the CMB temperature map with the SPOrt
$\Pi$ beam considering the instrument aligned to the
standard frame.
The result is shown in
Figure~\ref{contMapFig}. A typical
$E$-mode pattern is visible, with leading structures along
parallels and meridians for $Q$ and along 45$^\circ$--135$^\circ$
directions for $U$.
%%%%%%%%%%%%%%%%%%%%%%%%%%%%%%%%%%%%%%%%%%%%%%%%%%%%%%%%%%%%%%%%%%%%%%%%
\begin{figure}
 \resizebox{\hsize}{!}
   {\includegraphics[width = 1.0\hsize, angle = 0]{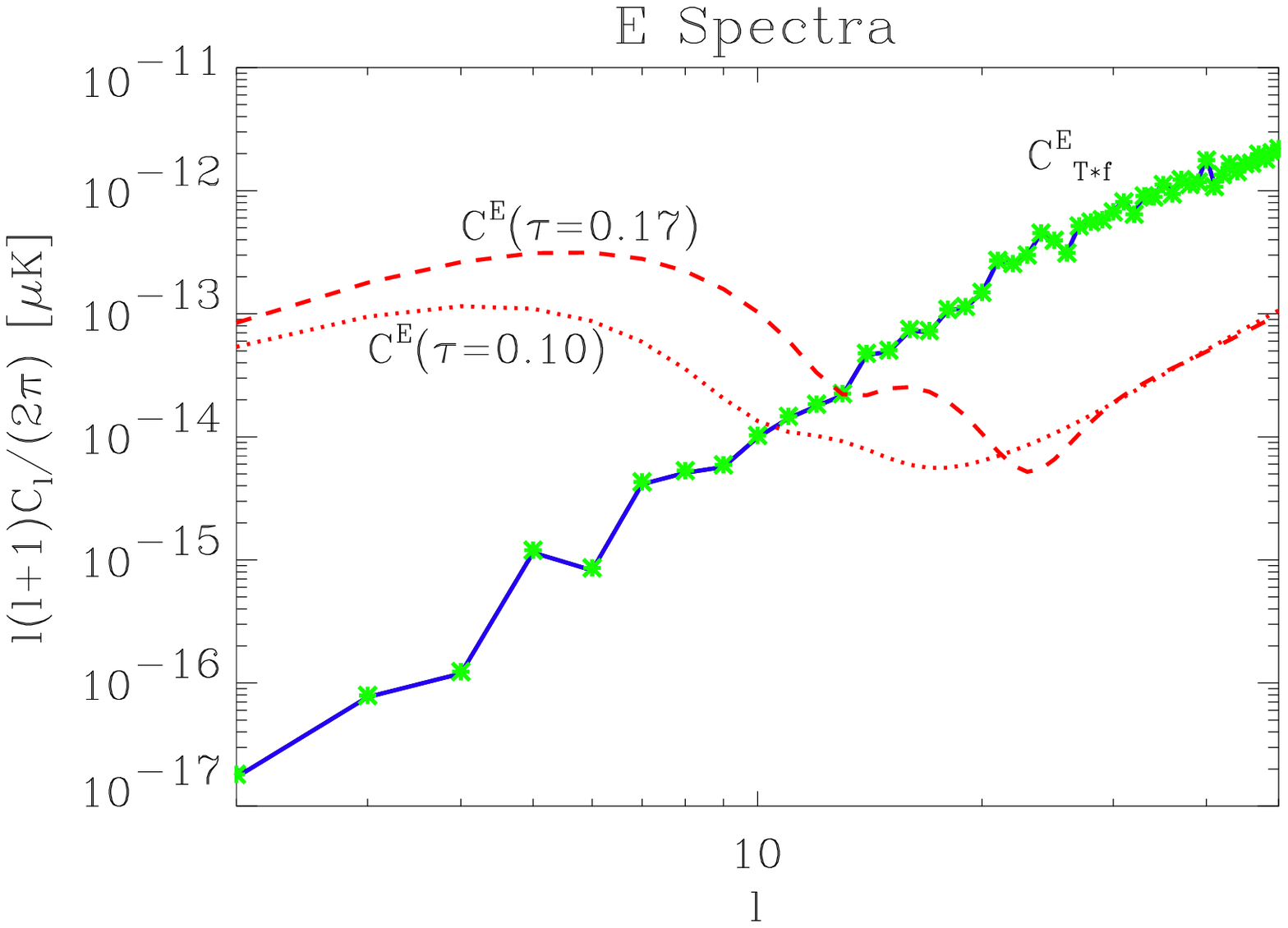}}
   {\includegraphics[width = 1.0\hsize, angle = 0]{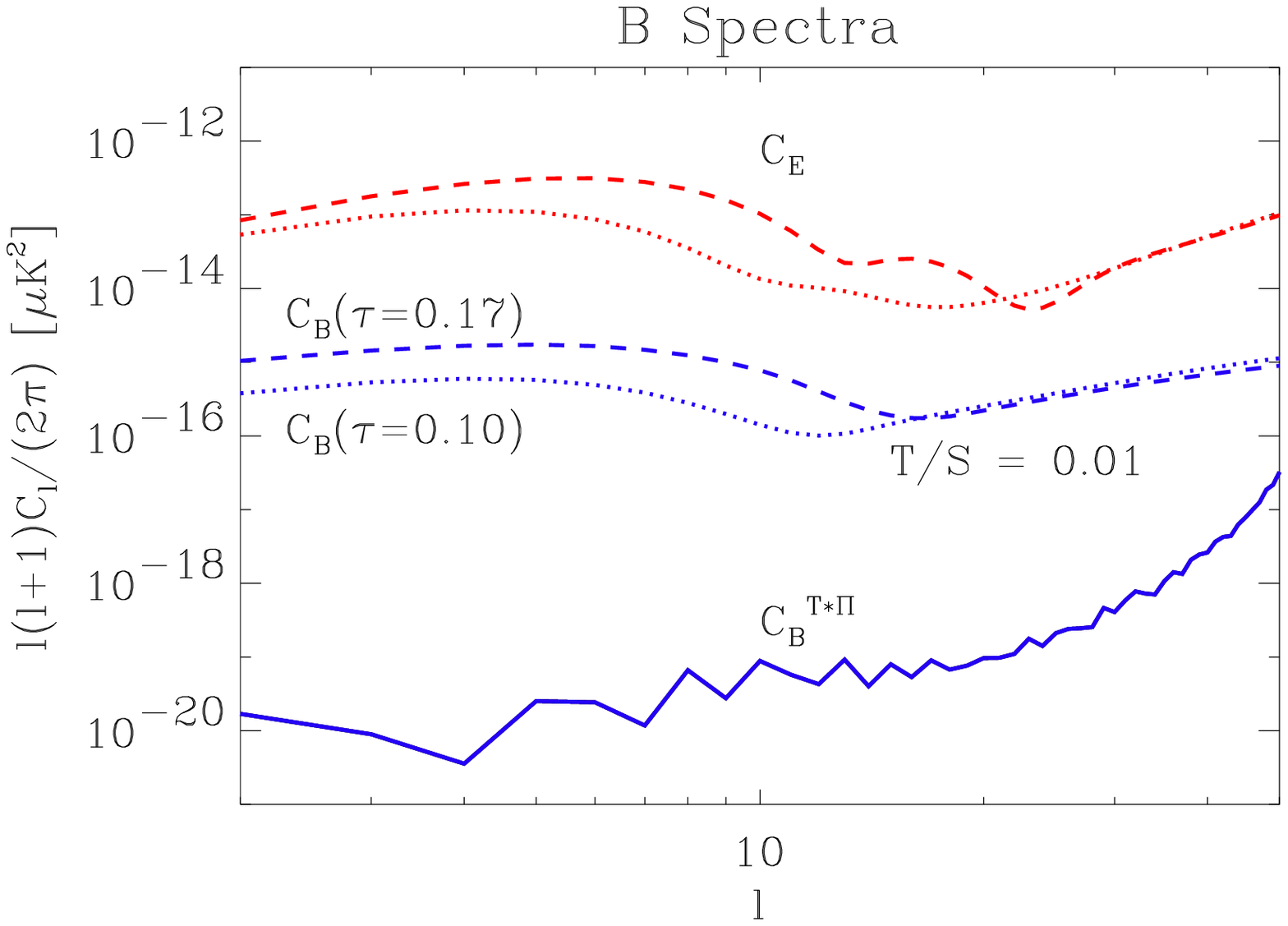}}
\caption{Top: $E$-mode power spectrum $C_E^{T\otimes \Pi}$
of the contamination maps of
Figure~\ref{contMapFig} (solid) together with the
product $C_T C_E^\Pi / 4\pi$ between the spectra of
the input temperature map and the instrumental beam $\Pi$ (stars).
The spectra are corrected for
the window function $W^2_\ell$ to account for
the smearing effects of the main beam.
For comparison, the CMBP $E$-mode spectrum
for the WMAP best-fit model with optical depth
$\tau = 0.17$ is reported (dashed).
A model with a smaller $\tau = 0.10$ is also shown (dotted).
Bottom: the same but for $B$. The spectrum of
 the contamination map $C_B^{T\otimes \Pi}$
(solid) is likely due to numerical noise.
 The expected CMBP spectra for two models with tensor-to-scalar ratio
 $T/S = 0.01$ are several orders of magnitude above this level.
 The corresponding $E$ spectra are also shown.}
\label{contPSFig}
\end{figure}
%%%%%%%%%%%%%%%%%%%%%%%%%%%%%%%%%%%%%%%%%%%%%%%%%%%%%%%%%%%%%%%%%%%%%%%%

The power spectra are shown in Figure~\ref{contPSFig}.
Here the spectra are corrected for the window function
of the main beam $W^2_\ell$
to account for its smearing effects on $E$ and $B$ spectra
(Zaldarriaga \& Seljak 1997).
As expected, the $B$-mode of the contamination map
is at a very low level, likely due to numerical precision, and,
in any case, is negligible when compared to the
faint level of the cosmological signal (here we use
a model with tensor-to-scalar ratio $T/S = 0.01$).

The most important contamination in the $E$-mode is
on FWHM scales and it rapidly decreases
on larger ones, as expected from the
$E$ spectrum of the SPOrt $\Pi$ pattern.
Moreover, it is well
fitted by the product between $C_T$ and $C_E^\Pi$.
Figure~\ref{contPSFig} shows its comparison with the
expected CMBP signal. At the large scale peak of CMBP
($\ell < 10$),
the instrumental polarization is nearly two orders
of magnitude below the spectrum of the signal.
This occurs not only for
the best fit WMAP model ($\tau = 0.17$), but also
for a less reionized Universe ($\tau=0.10$).
As a result, the SPOrt experiment does not look contaminated at
significant level on the most relevant angular scales
and no data cleaning seems to be needed.
In any case, should an experiment suffer from a relevant
leakage,
it would be possible to subtract the spurious contribution
estimated from both the Temperature Map and the $\Pi$ pattern.
Equation~(\ref{cecbeq}) together with the variance
equations for the Temperature spectra
(Zaldarriaga \& Seljak 1997) provides
a way to estimate the residual noise after the subtraction:
in the case of well known $\Pi$ pattern (negligible error)
and Temperature map with uniform
white noise, it results in
\begin{equation}
    \sigma_{C_E}(\ell) = {1 \over 4\pi} \, \sqrt{2  \over 2\ell+1}\,
                         {C_{E\ell}^\Pi  \over W^2_\ell}\,\,
                         {\omega^{-1}_T \over W^2_{T\ell}}
\end{equation}
with
$W^2_{T\ell}$ the window function of the Temperature map and
$\omega^{-1}_T = 4\pi\sigma^2_{T,{\rm px}}/N_{\rm px}$, where
$\sigma_{T,{\rm px}}$ is the pixel noise and $N_{\rm px}$ the number of 
pixels in the map. Finally,
the window function of the main beam $W^2_\ell$ accounts
for smearing effects on $E$.

\section{Conclusions}\label{conclusions}

We have derived the equations to compute
the instrumental polarization introduced by the antenna.
The contamination in both $Q$ and $U$
is the convolution of the unpolarized field
of the incoming radiation with an instrumental
polarization beam $\Pi = \Pi_Q + j \,\Pi_U$,
which is a
function of the co-
and cross-polar patterns of the antenna.
This result is general and independent of the
technique adopted to measure $Q$ and $U$. In particular,
it is valid for instruments either correlating the two
polarizations (either circular or linear)
or differentiating the two linear polarizations.

The special case of axisymmetric systems (like
circular dual-polarization feed horns
and on-axis mirrors)
presents special features:
\begin{itemize}
\item{} The instrumental polarization pattern $\Pi$ is axisymmetric
        in itself with intensity only depending  on the radial distance
        $\theta$ from the main axis;
\item{} Polarization angles have a radial
        pattern and are either along or perpendicular to
        the radial direction;
\item{} Both the instrumental beams $\Pi_Q$ and $\Pi_U$ of $Q$ and $U$
        change sign from quadrant
        to quadrant, which makes the contamination
        only dependent on the anisotropy of the radiation;
\item{} The contamination in the maps in the standard
        frame is independent of the instrument frame rotation and,
        in turn, of the scanning strategy of the experiment.
\end{itemize}
These features result in the relevant property
that the $\Pi$ pattern has no $B$-mode component,
leaving the $B$-mode of the sky signal uncontaminated.

This absence of leakage into the $B$-mode is an important result and
makes  axisymmetric systems
suitable solutions for the detection of
this faint signal.
In general, off-axis solutions do not satisfy the
axisymmetric condition
and an analysis has to be performed for each case
to evaluate the amount of instrumental polarization in $B$.

A further relevant result is that
the spectrum of the $E$ component of the instrumental polarization
is the product between the spectra of the unpolarized emission
map and the instrumental polarization pattern.
It peaks on FWHM scales, leaving significantly cleaner the larger ones.
This represents an important result for CMBP
experiments searching for signal on large scales,
where the very new information provided by CMBP reside. 
As an example, the contamination
generated in the SPOrt horns
at $\ell =2$ is four orders of magnitude lower than at
$\ell = 25$ (FWHM~$\sim 7^\circ$).
Moreover, at $\ell < 10$, where the CMBP has the large scale peak,
the instrumental contribution of the SPOrt antennae is
two orders of magnitude lower than the expected sky
signal, probably making not necessary any correction
for this systematic effect.

Experiments on small angular scales are in a different
situation. They look
for CMBP features close to the FWHM scale (the CMBP spectrum has
several {\it Doppler} peaks in the 5--30~arcmin range) where
the $E$-mode of the
$\Pi$ pattern peaks.
Moreover, the Doppler-peak pattern of the Temperature
will be reproduced in some way in the spectrum of the
contamination, leading to the possibility
to confuse the peak pattern of the CMBP $E$-mode.
A careful
analysis of the instrumental polarization generated by the antenna
system is thus mandatory for these experiments.

\begin{acknowledgements}
This work has been carried out in the frame of the SPOrt programme funded by
the Italian Space Agency (ASI). We thank the referee Jacques Delabrouille 
for useful comments. Some of the results in this paper have been derived
using the HEALPix (Gorski, Hivon \& Wandelt 1999).
We acknowledge the use of the CMBFAST package.
\end{acknowledgements}

\end{document}